\documentclass{PoS}

\usepackage{amsmath}
\usepackage{cite}

	\title{\vspace{10mm}Electromagnetic finite-size effects to the hadronic vacuum polarisation}

\ShortTitle{Electromagnetic finite-size effects to the hadronic vacuum polarisation}

\author{\speaker{Nils Hermansson-Truedsson}
\\
        Lund University\\
	E-mail: \email{nils.hermansson-truedsson@thep.lu.se}}

\author{Johan Bijnens\\
        Lund University\\
        E-mail: \email{bijnens@thep.lu.se}}

\author{James Harrison\\
        University of Southampton\\
        E-mail: \email{J.Harrison@soton.ac.uk}}

\author{Tadeusz Janowski\\
        University of Edinburgh\\
        E-mail: \email{t.janowski@ed.ac.uk}}

\author{Andreas J\"{u}ttner\\
        University of Southampton\\
        E-mail: \email{A.Juttner@soton.ac.uk}}

\author{Antonin Portelli\\
        University of Edinburgh\\
        E-mail: \email{antonin.portelli@ed.ac.uk}}

\abstract{
	In order to reach (sub-)per cent level precision in lattice calculations of the hadronic vacuum polarisation, isospin breaking corrections must be included. This requires introducing QED on the lattice, and the associated finite-size effects are potentially large due to the absence of a mass gap. This means that the finite-size effects scale as an inverse polynomial in $L$ rather than being exponentially suppressed. Considering the $\mathcal{O}(\alpha)$ corrected hadronic vacuum polarisation in QED$_{\mathrm{L}}$ with scalar QED as an effective theory, we show that the first possible term, which is of order $1/L^{2}$, vanishes identically so that the finite-size effects start at order $1/L^{3}$. This cancellation is understood from the neutrality of the currents involved, and we show that this cancellation is universal by also including form factors for the pions. We find good numerical agreement with lattice perturbation theory calculations, as well as, up to exponentially suppressed terms, scalar QED lattice simulations.
	
}

\FullConference{37th International Symposium on Lattice Field Theory - Lattice2019\\
		16-22 June 2019\\
		Wuhan, China}

\renewcommand{\logo}
{\raisebox{-5mm}
{\parbox{\textwidth}{\begin{flushright}
LU TP 19-47\\
October 2019\\
\end{flushright}
\includegraphics{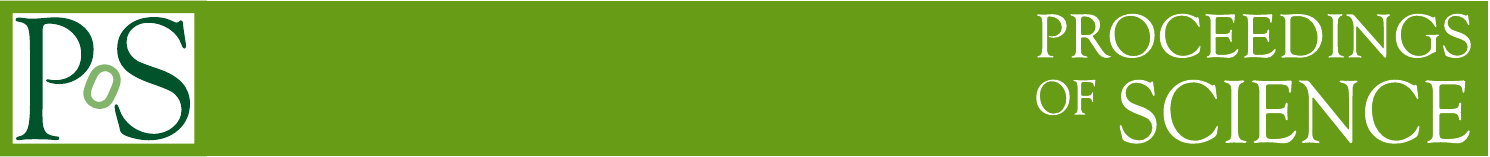}}
}}
\begin{document}

\section{Introduction}
Due to the current $3.5-4\sigma$ discrepancy between the predicted Standard Model (SM) value of the muon anomalous magnetic moment and the value measured experimentally at Brookhaven National Laboratory~\cite{Bennett:2006fi,Jegerlehner:2017lbd}
, there has been a lot of interest and effort to improve the precision on both the theoretical and experimental sides. The new experiment at Fermilab is expected to deliver new data with an error improved by a factor four within the coming year, and on the theory side a lot of work has gone into improving the precision on the contributions which currently have the largest uncertainty, i.e.~the hadronic ones. 

The hadronic vacuum polarisation (HVP) can be evaluated on the lattice 
and in order to reach (sub-)per cent precision isospin breaking effects have to be included. The electromagnetic corrections are particularly problematic due to the systematic effects coming from the finite-volume approximation on the lattice, and in order to provide useful input for the SM prediction these must be well under control. 

Including electromagnetism on the lattice requires handling the infrared singularities from the photons. We consider the HVP in the QED$_{\mathrm{L}}$ scheme, where all photon momentum modes of the form $k=(k_{0},\mathbf{0})$ are excluded as a means of regulating the mentioned singularities. This scheme has been used in e.g. Refs.~\cite{Hayakawa:2008an,Davoudi:2018qpl}.

The finite-size effects are then, for a lattice of geometry $T \times L^{3}$, given by an inverse polynomial in $L$ where the first a priori possible term is of order $1/L^{2}$. Using the methodology of Ref.~\cite{Davoudi:2018qpl} extended to 2-loop order for scalar QED (sQED) as an effective theory, we analytically show that the $1/L^{2}$ term identically vanishes, which can be understood from the neutrality of the currents entering into the HVP. By including form factors for the pions we show that this cancellation is universal, i.e.~independent of choosing sQED as the effective theory. We also check our result numerically, both by simulating sQED on the lattice as well as by comparing to lattice perturbation theory (LPT). A more detailed description of our calculation and also additional results not covered here can be found in Ref.~\cite{Bijnens:2019ejw}. 

\section{Some generalities}
The HVP is defined as the vector 2-point function according to
\begin{align}
	\Pi _{\mu\nu}(q)  =  \, \int d^{ 4}x \, e^{ iq\cdot x}\left\langle 0 \right| 
 \mathrm{T}[\, j_{\mu}(x)\, j_{\nu}^{ \, \dagger}(0)] \left| 0\right\rangle \, , 
\end{align}
where $j_{\mu} (x)$ is a vector current and $q^{ 2}$ denotes the Euclidean external photon momentum. From the Ward identity $q_{\mu}\Pi _{\mu \nu}(q)= 0 $ 
it follows that the HVP can be written as
\begin{align}
	\Pi _{\mu\nu}(q)= \left(q _\mu q_\nu-q^2 \delta 
_{\mu\nu}\right) \Pi (q^2) \, .
\end{align}
Note that the function $\Pi$ only depends on $q^{2}$, and it is its subtracted counterpart $\hat{\Pi}(q^{2})=\Pi (q^2) - \Pi (0) $ which is integrated over in the calculation of $a_{\mu}^{\textrm{HVP,LO}}$. We choose the photon momentum $q=(q_{0},\mathbf{0})$ so that $q^{2}=q_{0}^{2}$ and
\begin{align}
	\hat{\Pi} (q^{2}) = \frac{-1}{3q_{0}^{2}}\, \sum _{i=1}^{3} \Big( \Pi _{ii}(q)-\Pi_{ii}(0) \Big) \, .
\end{align}
The finite-size effects from electromagnetic (EM) corrections to the above quantity can be obtained using effective field theory techniques, see e.g.~Ref.~\cite{Davoudi:2018qpl}. As will be discussed in the next section, the EM finite-size effects can scale worse than those from pure QCD, and so we analytically derive the scaling with box size $L$ for the former. To do this, we use scalar QED (sQED) including only pions and photons, and calculate the $\mathcal{O}(\alpha)$ corrections to the HVP.

\begin{figure}[t!]
  \centerline{
  \includegraphics[width=0.75\linewidth]{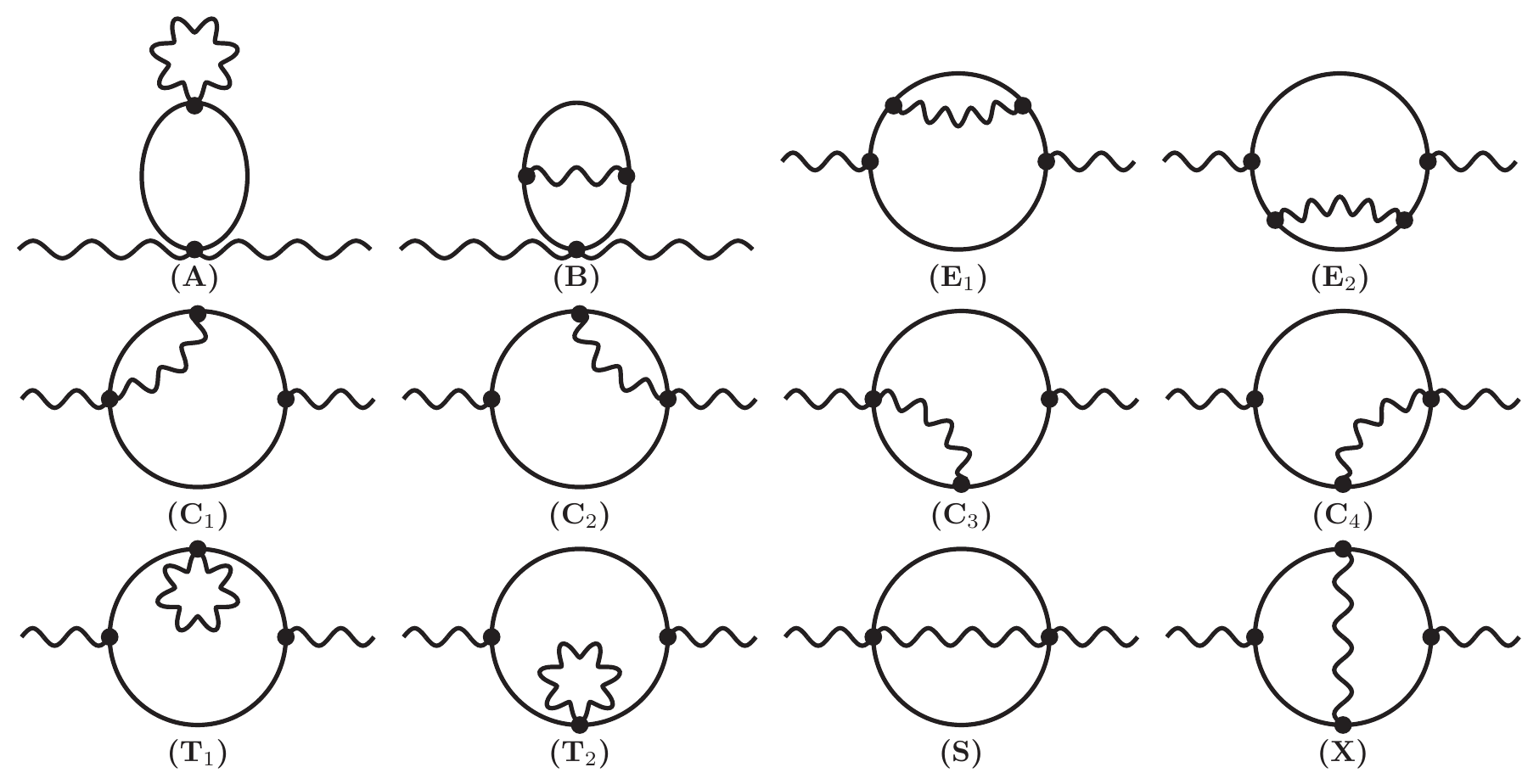}}
  \caption{The connected diagrams contributing at $\mathcal{O}(\alpha)$.}
  \label{fig:DiagramFig}
\end{figure}

In the following we only consider the connected diagrams\footnote{The disconnected diagrams at order $\alpha$ are in QED$_{\mathrm{L}}$ excluded for our choice of kinematics, i.e. for $\mathbf{q}=\mathbf{0}$.} at $\mathcal{O}(\alpha)$. These are shown in fig.~\ref{fig:DiagramFig}. In the subtraction of $q^{2}=0$, diagrams (A) and (B) automatically vanish. In effect, only five diagram topologies remain, and the HVP can at order $\alpha$ be written as a weighted sum over the different contributions according to
\begin{align}\label{eq:totalfveff}
	\hat{\Pi} \left( q^{2}\right)  \stackrel{\mathcal{O}(\alpha)}{=}\,  2\, \hat{\Pi}_{E} (q^{2})+ 4\, \hat{\Pi}_{C}  (q^{2}) + 2\, \hat{\Pi}_{T} (q^{2})+\hat{\Pi}_{S} (q^{2}) + \hat{\Pi}_{X} (q^{2})= \sum _{U} a_{U}\, \hat{\Pi} _{U} (q^{2})\, ,
\end{align}
where each diagram (U) contributes through $\hat{\Pi }_{U}(q^2)$. The numerical factors $a_{U}$ are easily seen in fig.~\ref{fig:DiagramFig}. In the following, we therefore only need to consider (E), (C), (T), (S) and (X).

\section{Finite-size effects}
Due to the mass gap in QCD, pions produce exponentially suppressed finite-size effects. However, without such a gap in QED the finite-size effects from EM corrections can scale as an inverse polynomial in the lattice size $L$. Below, the first non-vanishing term in this polynomial for the $\mathcal{O}(\alpha)$ corrected HVP is derived within sQED. It is assumed in the analytic derivation that the time extent $T\rightarrow \infty$. Our approach is a generalisation of the procedure for 1-loop integrals in Ref.~\cite{Davoudi:2018qpl}. 

Each diagram (U) has the form
\begin{align}
	\hat{\Pi} _{U}(q^{2}_{0}) =  \int \frac{d^{4} k}{(2\pi)^4} \frac{d^4 \ell}{(2\pi)^4}\, \hat{\pi} _{U}\left( k,\ell ,q_{0}\right) \, ,
\end{align}
for photon and pion loop momenta $k$ and $\ell$, respectively. Here, $\hat{\pi} _{U}\left( k,\ell ,q_{0}\right) $ is the subtracted loop integrand of diagram (U). The first step in the calculation is to compute the two energy integrals using contour integration, i.e. the $k_{0}$ and $\ell _{0}$ integrals. One then obtains
\begin{align}
	\hat{\rho } _{U}\left( \mathbf{k},\pmb{\ell},q_{0}\right) =  \int \frac{d k_{0}}{2\pi} \frac{d \ell_{0}}{2\pi}\, \hat{\pi } _{U}\left( k,\ell ,q_{0}\right) \, .
\end{align}
From this one can access the finite-size effects by considering the sum integral difference
\begin{align}
	\Delta \hat{\Pi} _{U}\left( q_{0}^{2}\right) = \left( \frac{1}{L^{6}}\left. \sum_{\mathbf{k}} \right.  ^{\prime }\sum _{\pmb{\ell}}-\int \frac{d^{\, 3}\mathbf{k}}{\left( 2\pi\right) ^{3}}\frac{d^{\, 3}\pmb{\ell}}{\left( 2\pi\right) ^{3}}\right)  \hat{\rho} _{U}\left( \mathbf{k},\pmb{\ell},q_{0}\right) \,,
\end{align}
where the finite-volume sums are over discretised momenta of the form
$\mathbf{k}=\frac{2\pi}{L}\mathbf{n}$ for a vector $\mathbf{n}$ of integers. The primed sum indicates the choice of $\mathrm{QED}_{\mathrm{L}}$, i.e.~the exclusion of all $\mathbf{k}=\mathbf{0}$ modes. Note further that since $q^2>0$, only photon lines can go on-shell and produce terms in inverse powers of $L$.
The pion sum-integral difference can be written as an integral up to exponentially suppressed terms, i.e. by using the Poisson summation formula. The finite-size corrections then take the form
\begin{align}
	\Delta \hat{\Pi } _{U}\left( q_{0}^{2}\right) = \left( \frac{1}{L^{3}}\left. \sum_{\mathbf{k}} \right.  ^{\prime }-\int \frac{d^{ 3}\mathbf{k}}{\left( 2\pi\right) ^{3}}\right)  \int \frac{d^{ 3}\pmb{\ell}}{\left( 2\pi\right) ^{3}} \, \hat{\rho} _{U}\left( \mathbf{k},\pmb{\ell},q_{0}\right) +\mathcal{O}\left( e^{-m_{\pi}L}\right) \,. 
\end{align}
From this expression one may next isolate the singular terms in $\mathbf{k}$ and Taylor expand these in $1/L$. Denoting the terms multiplying $(2\pi/|\mathbf{k}|)^{j}$ in this expansion as $u_{j}\left( \hat{\mathbf{n}},\pmb{\ell} ,q_{0}\right) $ results in   
\begin{align}\label{eq:fveffformula}
	\Delta \hat{\Pi} _{U}\left( q_{0}^{2}\right) = \frac{\xi _{1}^{U}\left( q_{0}^{2}\right) }{L^{2}}+\frac{\xi _{0}^{U}\left( q_{0}^{2}\right) }{L^{3}}+\mathcal{O}\left(\frac{1}{L^4},e^{\, -m_{\pi}L}\right) \, ,
\end{align}  
where the coefficients $\xi _{j}^{U} \left( q_{0}^{2} \right) $ are defined through 
\begin{align}
	\xi _{j}^{U} \left( q_{0}^{2}\right) =\Delta  _{\mathbf{n}}^{\prime } \left[ \frac{1}{\left| \mathbf{n}\right| ^{ j}}\int \frac{d^{ 3}\pmb{\ell}}{\left( 2\pi\right) ^{3}}\, u_{j}\left( \hat{\mathbf{n}},\pmb{\ell} ,q_{0}\right) \right]\, .
\end{align} 
Here $\Delta  _{\mathbf{n}}^{\prime }$ is the sum-integral difference operator
\begin{align}
	\Delta  _{\mathbf{n}}^{\prime }=\left.\sum_{\mathbf{n}}\right.^{\prime }-\int d^{\, 3}\mathbf{n}\, , 
\end{align}
i.e. the same as in Ref.~\cite{Davoudi:2018qpl} and depends on the choice of QED$_{\mathrm{L}}$. The coefficients $\xi _{j}^{U} \left( q_{0}^{2} \right) $ depend on a set of numbers commonly denoted as $c_{j} =\Delta  _{\mathbf{n}}^{\prime }|\mathbf{n}|^{-j}$, where the first two are $c_{0}=-1$ and $c_{1}=-2.83729748$~\cite{Davoudi:2018qpl}, as well as on a set of dimensionless integrals
\begin{equation}
\Omega_{\alpha,\beta}(z)=\frac{1}{2\pi^2}\int_0^{\infty}d x\,
  x^2 \frac{1}{( x ^{2}  +1)^{\frac{\alpha}{2}}[z+4(x ^{2}+1)]^{\beta}} \, , 
\end{equation}
where $z=q_0^2/m_{\pi}^2$. These integrals are convergent for $\alpha+2\beta >3$ and in that case very easy to calculate numerically.

It is therefore possible to access the finite-size effects for each diagram (U) through~(\ref{eq:fveffformula}), and add them all as in~(\ref{eq:totalfveff}) to obtain the total electromagnetic finite-size effects to the HVP at order $\alpha$.

\subsection{Analytic results}
To avoid lengthy formulae, only the final result for $\Delta \hat{\Pi}(q^{2})$ is given below. It should, however, be noted that each of the five topologies (E), (C), (T), (S) and (X) contributes with at least a term of order $1/L^{2}$. Adding these results according to~(\ref{eq:totalfveff}) and suppressing the dependence on $z$ for $\Omega _{\alpha , \beta}(z)$ then yields
\begin{align}\label{eq:finalhvp}
\Delta \hat{\Pi} (q^2) =
	\frac{c_0}{m_{\pi}^3 L^3}
\Bigg(
           \frac{16}{3}\Omega_{0,3}
          + \frac{5}{3}\Omega_{2,2}
          - \frac{40}{9}\Omega_{2,3}
          + \frac{3}{8}\Omega_{4,1}
          - \frac{7}{6}\Omega_{4,2}
          - \frac{8}{9}\Omega_{4,3}
\Bigg)\, . 
\end{align}
As can be seen, the individual terms of order $1/L^{2}$ cancel identically in the sum of all diagrams. This can be understood from the neutrality of the currents involved in the vector 2-point function, since the photons probing large distances and hence the finite-size effects cannot resolve the charge separation between the pions in the current. In fact, using charged currents by also including the $\pi ^{0}$ does not yield a cancellation at order $1/L^{2}$, as expected.  

It should be noted that sQED, i.e. with point-like pions, was used to obtain the result in~(\ref{eq:finalhvp}). However, by including form factors it is possible to show that the cancellation in fact is universal so that the form factors only change the overall coefficient of the $1/L^{3}$ term.

The implications of our result is that for reasonably sized $m_{\pi}L$ the finite-size effects are negligible. To see this, note that for $m_{\pi}L\geq4$ one has $1/(m_{\pi}L)^3\leq 1.5\%$, i.e.~the finite-size effects correspond to per cent level corrections to the $\mathcal{O}(\alpha)\sim 1\%$ corrected HVP. Thus, for the currently sought precision, these finite-size effects are negligible for reasonable $m_{\pi}L$ as long as the overall coefficient multiplying the cubic term is not unnaturally large. This is of course something that will have to be checked in the full QCD+QED theory.

\section{Numerical validation}
\begin{figure}[t!]
  \centerline{
  \includegraphics[width=0.45\linewidth]{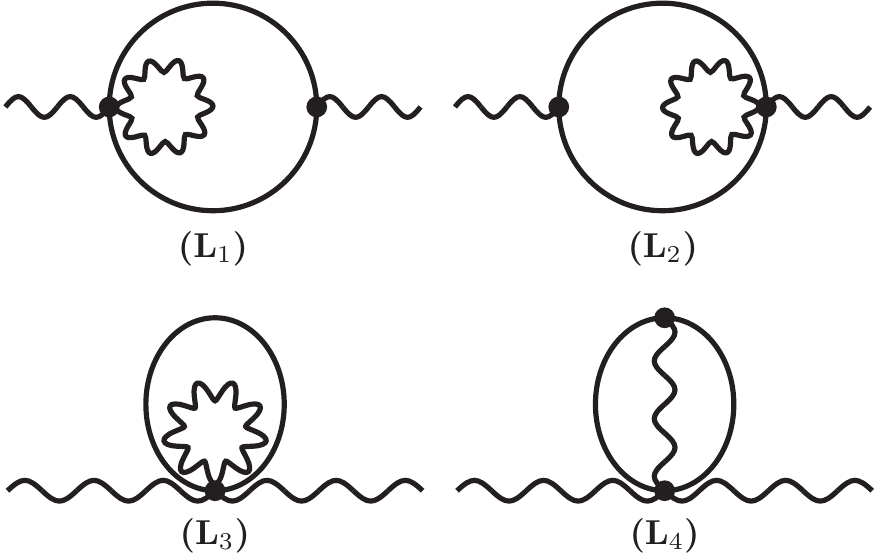}}
  \caption{The additional diagrams in the lattice theory contributing at $\mathcal{O}(\alpha)$.}
  \label{fig:LattDiagrams}
\end{figure}

In order to check the validity of the analytic results presented in the previous section, we also numerically calculate the finite-size scaling by on the one hand simulating sQED on lattices for various $L$ and on the other by calculating the 2-loop integrals for all diagram topologies in sQED LPT, also at several volumes. We will here not go into the details of these numerical calculations, as they can be found in Ref.~\cite{Bijnens:2019ejw}. Instead, we focus on the implications of the comparison to the analytic results. It is important to note that in the discretised theory four additional diagrams appear, see fig.~\ref{fig:LattDiagrams}, which all vanish in the continuum limit, i.e.~when the lattice spacing $a\rightarrow 0$. However, diagrams (L$_{3,4}$) automatically vanish in the subtraction of $q^{2}=0$ in $\hat{\Pi}(q^{2})$ and so need not be considered.

On the lattice the pions are by definition in finite volume (FV), but in LPT it is possible to allow the pions to be in infinite volume (IV), just as in our analytic approach. By thus calculating the LPT loop integrals for both FV and IV pions allows to validate both the lattice simulations as well as the use of the Poisson summation formula for the pions in the analytic derivation.

\begin{figure}[t!]
  \centerline{
  \includegraphics[width=0.75\linewidth]{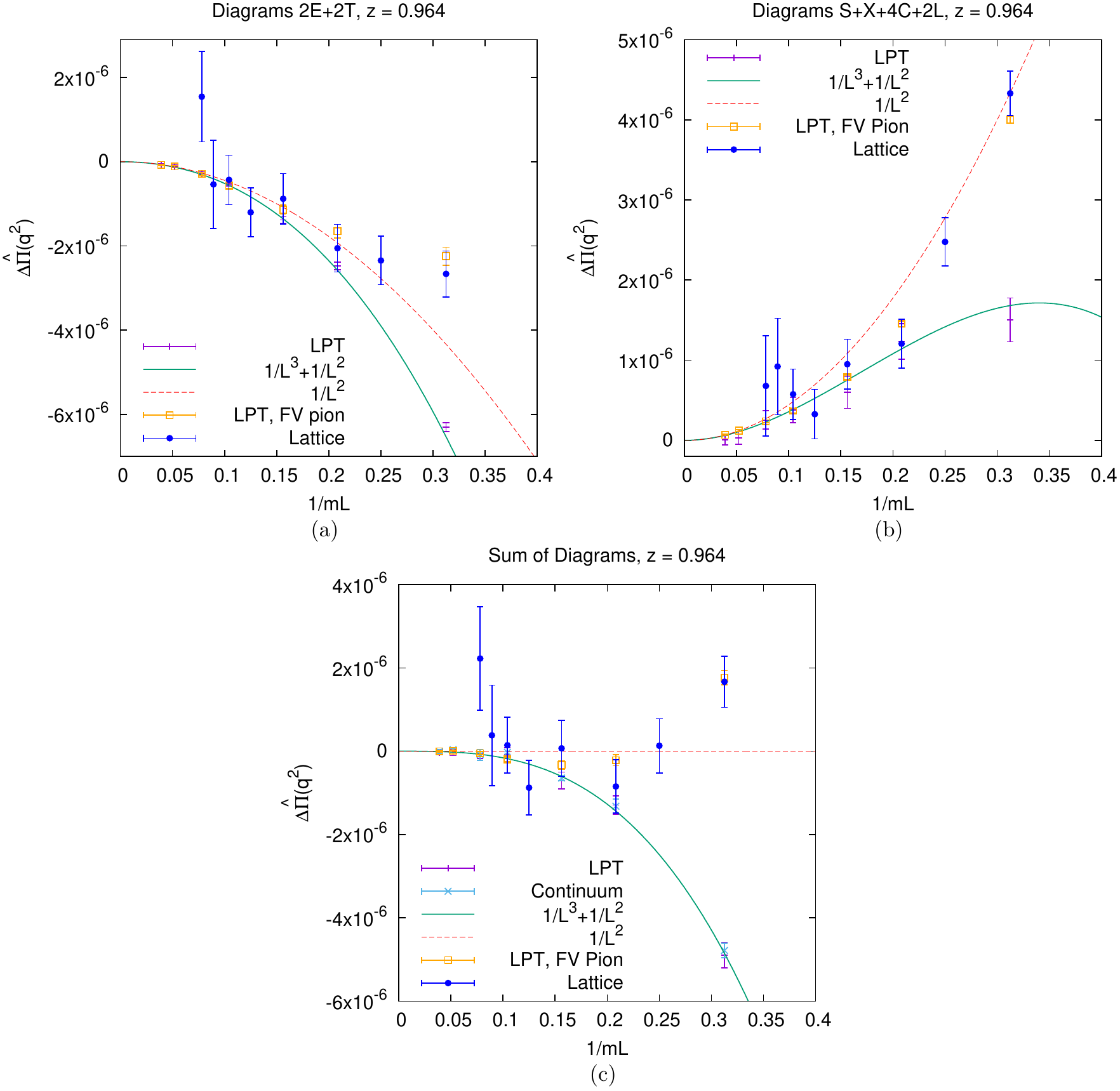}}
  \caption{The finite-size effects $\Delta \hat{\Pi}(q^{2})$ from both the analytical and numerical calculations for the diagram combinations (a) $2\cdot\textrm{(E)}+2\cdot\textrm{(T)}$, (b) $\textrm{(S)}+\textrm{(X)}+4\cdot \textrm{(C)}+2\cdot \textrm{(L)}$ and (c) the full set of diagrams.}
  \label{fig:FinalResults}
\end{figure}

We choose $am_{\pi}=0.2$ and $aq_{0}= 8\pi/128$ so that $z\approx 0.964$, and do the lattice calculations at eight different volumes between $L/a = 16$ and $L/a = 64$. The finite-size effects are shown for the specific diagram combinations\footnote{For these combinations there is no cancellation of the $1/(m_{\pi}L)^{2}$ terms.} $2\cdot\textrm{(E)}+2\cdot\textrm{(T)}$ and $\textrm{(S)}+\textrm{(X)}+4\cdot \textrm{(C)}+2\cdot \textrm{(L)}$ in figs.~\ref{fig:FinalResults}(a)--(b) as well as for the full sum of diagrams in fig.~\ref{fig:FinalResults}(c). The red dashed line corresponds to only the $1/(m_{\pi}L)^{2}$ term in the analytic formula whereas the solid green line corresponds to the full analytic result including also the $1/(m_{\pi}L)^{3}$ term. The dark blue points are from the lattice calculations and the orange points are from LPT for FV pions. Finally, the purple points are LPT data for IV pions and the continuum extrapolation of these are shown in light blue. First of all note that the IV pion LPT data agrees excellently with the full analytic results, and that the FV pion LPT points agree nicely with the lattice data. For $m_{\pi}L\geq 4$ there is good agreement between all four approaches, but for smaller $m_{ \pi}L$ one starts to see a deviation between the IV pion case and FV pion case. These differences can therefore be concluded to arise in the use of the Poisson summation formula, i.e. the neglected exponentially suppressed terms are of considerable size for such $m_{\pi}L$.  

\section{Conclusions}
We derive the analytic scaling in lattice size $L$ of the first electromagnetic corrections to the HVP. This is done in QED$_{\mathrm{L}}$ using scalar QED as an effective theory. We show that the first possible term of order $1/L^{2}$ identically vanishes, which can be understood from the neutrality of the currents involved, so that the first term starts at order $1/L^{3}$. We also show that this cancellation is universal, i.e.~independent of choosing sQED to derive the finite-size effects. The implication of our results is that for reasonably sized values of pion masses and $L$, $m_{\pi}L\geq 4$, say, the finite-size effects are negligible for the currently sought precision. This should also be checked for full QCD+QED simulations, since the final aim is to reduce the error on the SM prediction of the muon $g-2$.

\section*{Acknowledgments}
    Lattice computations presented in this work have been
    performed on DiRAC equipment which is part of the UK National
    E-Infrastructure, and on the IRIDIS High Performance Computing Facility at
    the University of Southampton. T.J. and A.P. are supported in part by UK
    STFC grants ST/L000458/1 and ST/P000630/1. A.P. also received funding from
    the European Research Council (ERC) under the European Union's Horizon 2020
    research and innovation programme under grant agreement No 757646. J.B. and
    N.H.T. are supported in part by the Swedish Research Council grants contract
    numbers 2015-04089 and 2016-05996, and by the European Research Council
    (ERC) under the European Union's Horizon 2020 research and innovation
    programme under grant agreement No 668679. J.H. was supported by the EPSRC
    Centre for Doctoral Training in Next Generation Computational Modelling
    grant EP/L015382/1. A.J. received funding from STFC consolidated grant
    ST/P000711/1 and from the European Research Council under the European
    Union's Seventh Framework Program (FP7/2007- 2013) / ERC Grant agreement
    279757. 

\bibliographystyle{JHEP}
\bibliography{refs}

\end{document}